\documentclass[11pt]{article}

\usepackage[final]{acl}

\usepackage{times}
\usepackage{latexsym}
\usepackage{pgffor}
\usepackage{subcaption}
\usepackage[table]{xcolor}
\usepackage{colortbl}
\usepackage[T1]{fontenc}

\usepackage[utf8]{inputenc}

\usepackage{microtype}

\usepackage{inconsolata}

\usepackage{graphicx}
\usepackage{ragged2e}
\usepackage{array}
\usepackage{tabularx}
\usepackage{multirow}
\newcommand{\dataset}{{\em Sociocultural Statements}}

\title{Understanding Cultural Alignment in Multilingual LLMs \\via Natural Debate Statements}

\author{Vlad-Andrei Negru$^{1}$, Camelia Lemnaru$^{1}$,  \textbf{Mihai Surdeanu$^{2}$, Rodica Potolea$^{1}$ }\\\\
$^1$Department of Computer Science, Technical University of Cluj-Napoca, Cluj-Napoca, Romania  \\
	$^2$Department of Computer Science, University of Arizona, Tucson, USA \\ 
		\{vlad.negru, camelia.lemnaru, rodica.potolea\}@cs.utcluj.ro, \\\ msurdeanu@arizona.edu  \\
}

\definecolor{lightgreen}{RGB}{200,245,200}
\newcolumntype{Y}{>{\centering\arraybackslash}X}
\newcolumntype{F}[1]{%
    >{\raggedright\arraybackslash\hspace{0pt}}p{#1}}%

\newcolumntype{Z}{>{\raggedleft\arraybackslash}X}

\newcolumntype{J}{>{\justifying\setlength{\parindent}{0pt}\arraybackslash}X}

\begin{document}
\maketitle
\begin{abstract}

In this work we investigate the sociocultural values learned by large language models (LLMs). We introduce a novel open-access dataset, {\dataset}, constructed from natural debate statements using a multi-step methodology. The dataset is synthetically labeled to enable the quantization of sociocultural norms and beliefs that LLMs exhibit in their responses to these statements, according to the Hofstede cultural dimensions. 
We verify the accuracy of synthetic labels using human quality control on a representative sample.
We conduct a comparative analysis between two groups of LLMs developed in different countries (U.S. and China), and use as a comparative baseline patterns observed in human measurements. Using this new dataset and the analysis above, we found that {\em culturally-distinct LLMs reflect the values and norms of the countries in which they were developed}, highlighting their inability to adapt to the sociocultural backgrounds of their users.

\textcolor{red}{Note: We use these cultures to identify sociocultural biases in LLMs. It is not our intention to compare cultures. }

\end{abstract}

\section{Introduction}









Over the past few years, rapid advancements have pushed the performance boundaries of large language models (LLMs), driven by large-scale pretraining and post-training alignment methods, such as instruction tuning and reinforcement learning  from human feedback. These models are widely used by hundreds of millions of people 
for a variety of tasks
\cite{chatgptUse}.  


However, as noted in prior work \cite{johnson2022ghostmachineamericanaccent, shen-etal-2024-understanding, AIWeird, 10.1162/COLI.a.14}, general-purpose AI systems often fail to adapt to the diverse sociocultural backgrounds and value systems of their users. This creates the risk of imposing a set of ostensibly universal beliefs, potentially shaping and distorting the sociocultural values and norms of their users \cite{Jakesch_2023, chawla-etal-2023-social}. Despite existing research advocating for multicultural annotation practices \cite{plank-2022-problem, prabhakaran2022culturalincongruenciesartificialintelligence}, the development of LLMs continues to rely predominantly on a single alignment paradigm.


\begin{figure*}[t!]
    \centering
    \includegraphics[width=1\textwidth]{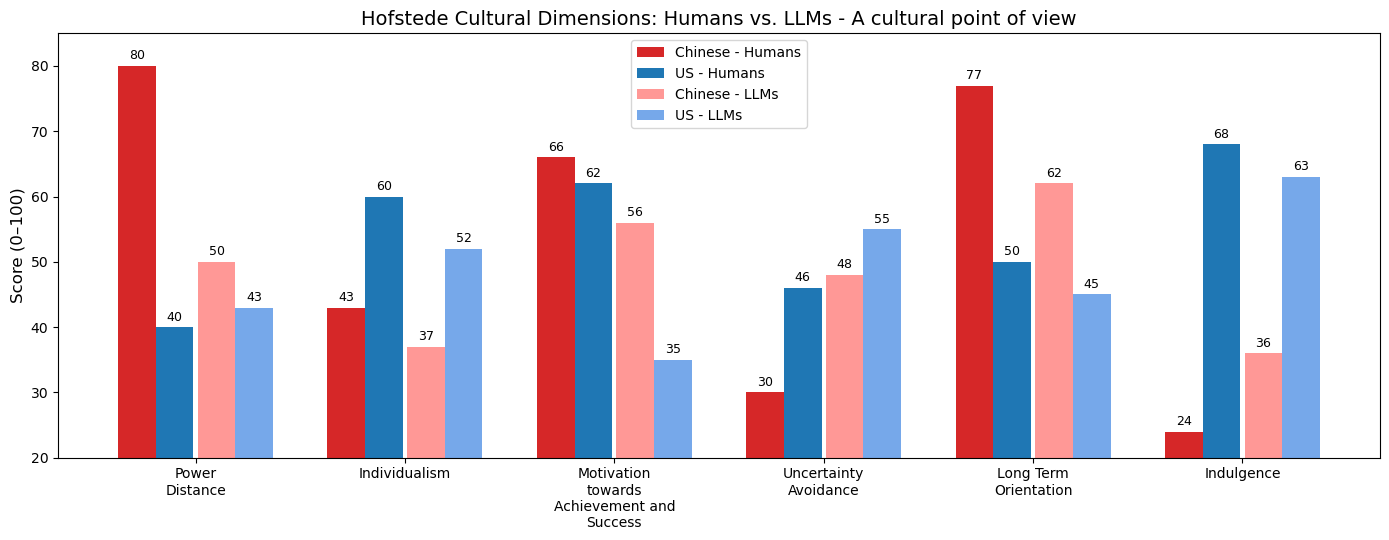}
    \vspace{-5mm}
    \caption{Scoring of humans and LLMs according to the Hofstede dimensions. LLMs were prompted to agree or disagree with a given statement, and each response was assigned a polarity score ranging from $-2$ to $+2$ along the corresponding Hofstede dimension. These scores were then averaged and linearly mapped to a 0--100 scale to ensure comparability with human measurements. We observe that the LLMs largely follow the trends observed in human measurements across all six dimensions.}
    \vspace{-4mm}
    \label{fig:english_vs_chinese}
\end{figure*}

Here, we quantify the extent to which {\em LLM responses reflect the sociocultural norms of the countries in which they were developed}. Our analysis focuses on LLMs originating from two of the most influential countries in current LLM development: the United States and China. In particular, the contributions of our paper are:

{\flushleft {\bf (1)}} We introduce {\dataset}, a dataset obtained from natural debate statements through a multi-stage methodology. The statements are synthetically labeled to enable the quantification of sociocultural values in LLM responses, grounded in Hofstede’s cultural dimensions \cite{Hofstede:1991}. 
We verify the quality of these annotations using human quality control on a representative sample.
To our knowledge, \dataset\ is substantially larger than prior datasets in this area, containing 281 statements that were voted as meaningful by users of Kialo, an online debate platform.\footnote{\label{kialourl}\url{https://www.kialo.com}} The dataset is publicly available (see \autoref{sec:dataset_appendix}).

{\flushleft {\bf (2)}} 
Using our novel dataset, we assess the sociocultural values expressed by multiple LLMs developed in two countries (the U.S. and China) under a multilingual prompting setup (English and Chinese). We perform comparative analyses across multiple levels of granularity: (a) a comparison of aggregated responses at the sociocultural group level; (b) an examination of sociocultural biases at the individual model level; and (c) an analysis of how the prompting language influences the sociocultural values expressed by each model.

{\flushleft {\bf (3)}} We compare our results against a baseline derived from human measurements, and highlight important observations, grounded in social science theory. Our main observation, as illustrated in \autoref{fig:english_vs_chinese}, is that culturally-distinct LLM groups follow trends similar to those observed among people from the same culture, across all Hofstede dimensions. Moreover, the sociocultural values seen in LLM responses are resilient to language switches.

\section{Related work}

Motivated by the view that language is fundamentally embedded in social and cultural context \cite{yang-etal-2025-socially, shen-etal-2024-understanding}, recent works try to examine whether LLMs encode sociocultural values as latent properties of model behavior. Analyses of free-form generations in culturally salient scenarios indicate that although LLMs exhibit culturally patterned tendencies, these signals appear to be weak, context-dependent, and internally inconsistent, often defaulting to dominant norms rather than reflecting stable cultural value representations. 

Several works ground cultural evaluation in established social science frameworks, most notably Hofstede’s cultural dimensions \cite{Hofstede:1991}. \citet{kharchenko2025llmsrepresentvaluescultures} introduce a large-scale multilingual advice benchmark covering five Hofstede dimensions and 36 country–language pairs, combining manual and LLM-generated prompts, persona conditioning, multilingual translation, and mixed automatic–manual annotation. Although LLMs distinguish opposing sociocultural values, they fail to apply them consistently in advice, even under explicit country personas, exposing a gap between conceptual recognition and behavioral expression.

Using a standardized survey-based approach, \citet{masoud-etal-2025-cultural} operationalize the Hofstede VSM13 questionnaire to measure cultural alignment in LLMs. By varying persona prompts, decoding parameters, and language-specific fine-tuning, they show that cultural alignment is weak and unstable across models; while GPT-4 performs best, results are highly sensitive to prompt and sampling choices, indicating fragile rather than robust cultural representations.

Finally, \citet{ma2025openopinionembracingopenendedness} argue that closed-ended evaluations underestimate LLMs’ social reasoning. Drawing on survey methodology and qualitative social science, they advocate for open-ended generation to better capture interpretability and heterogeneity in social simulation, while highlighting open challenges in benchmarking, evaluation reliability, and ethical risks posed by plausible but unverifiable outputs.

Complementing these research efforts, practitioners have also conducted informal experiments to assess AI’s cultural competence: the Culture Factor Group\footnote{\label{culturefactor}\url{https://www.theculturefactor.com/}} conducts an exploratory cultural audit by manually designing a 42-question instrument aligned with Hofstede’s 6 dimensions and prompting several LLMs (e.g. GPT-4, Gemini) with explicit national personas, with and without theory priming, revealing substantial inconsistencies between expected and model-generated cultural profiles \cite{culturefactor2024aimeetsculture}.

Most prior work probes cultural alignment by explicitly injecting cultural identity through prompts (e.g., nationality personas or culture-specific scenarios), which conflates cultural knowledge elicitation with genuine cultural representation and limits insights into how such values are internally encoded. 
Others use synthetic questions that are directly aligned with Hofstede dimensions but are unlikely to occur in natural dialogues.
In contrast, we assess sociocultural values {\em implicitly} by analyzing how LLMs evaluate the truthfulness of {\em natural} statements.
We align these evaluations with Hofstede’s cultural dimensions and validate the resulting mappings through human quality control. This indirect assessment allows us to probe the representations of sociocultural values through statements that are more likely to be similar to naturally-occurring LLM interactions, without explicitly biasing the model towards certain social norms. Moreover, our methodology produces a significantly larger set of questions/statements for probing LLMs than previous works. 

\section{The Hofstede cultural dimensions}
\label{sec:hofstede}

Our analysis relies on the Hofstede cultural dimensions framework \cite{Hofstede:1991}. For completeness, we provide brief definitions of its dimensions in \autoref{tab:hofstede_dimensions}, also specifying the traits of each polarity (i.e., negative and positive).

\begin{table*}[ht!]
\begin{center}
\resizebox{1.0\textwidth
}{!}{
\begin{tabularx}{1.4\textwidth}{ p{5cm} p{9cm} J J }
\hline
\textbf{Dimension} & \textbf{Definition} & \textbf{Positive polarity} & \textbf{Negative polarity}\\
\hline
Power Distance Index (PDI) & Evaluates the degree to which citizens tolerate unequal power relationships between various groups. & acceptance of power inequality & equalitarian values \\
\hline
Collectivism vs. Individualism (IDV) & Evaluates the degree to which a country organizes people into homogeneous groups that have specific duties and obligations within the social structure. & individualistic pursuit & greater good of groups \\
\hline
Uncertainty Avoidance (UAI) & Evaluates the extent to which the members of a culture feel threatened by ambiguous or unknown situations. & high levels of risk aversion & risks acceptance \\
\hline
Motivation towards Achievement and Success (MAS) & Formerly Masculinity vs. Femininity, it evaluates societal tendencies regarding gender roles, gender equality, and traditional gender values. & goal-oriented behavior, wealth accumulation, clearly defined gender roles. & collective quality of life, and flexible gender roles \\
\hline
Short-term vs. Long-term Orientation (LTO) & Evaluates the degree to which a society acts to further short-term objectives at the expense of long-term goals and vice versa. & immediate gratification & sacrifice of instant results for the greater long-term good \\
\hline
Indulgence vs. Restraint (IND) & Evaluates the degree to which a society allows free gratification of basic and natural human desires related to enjoying life, as opposed to restraining through strict social norms and regulations. & life enjoyment, entertainment, luxuries, and showy displays of
socioeconomic status. & importance of modesty, suppression, and self-control. \\
\hline
\end{tabularx}
}
\end{center}
\vspace{-3mm}
\caption{Hofstede dimensions definitions together with interpretations of response polarities.}
\vspace{-3mm}
\label{tab:hofstede_dimensions}
\end{table*}

\section{Dataset construction and alignment}

\begin{figure}[t!] 
    \centering
    \includegraphics[width=0.90\linewidth]{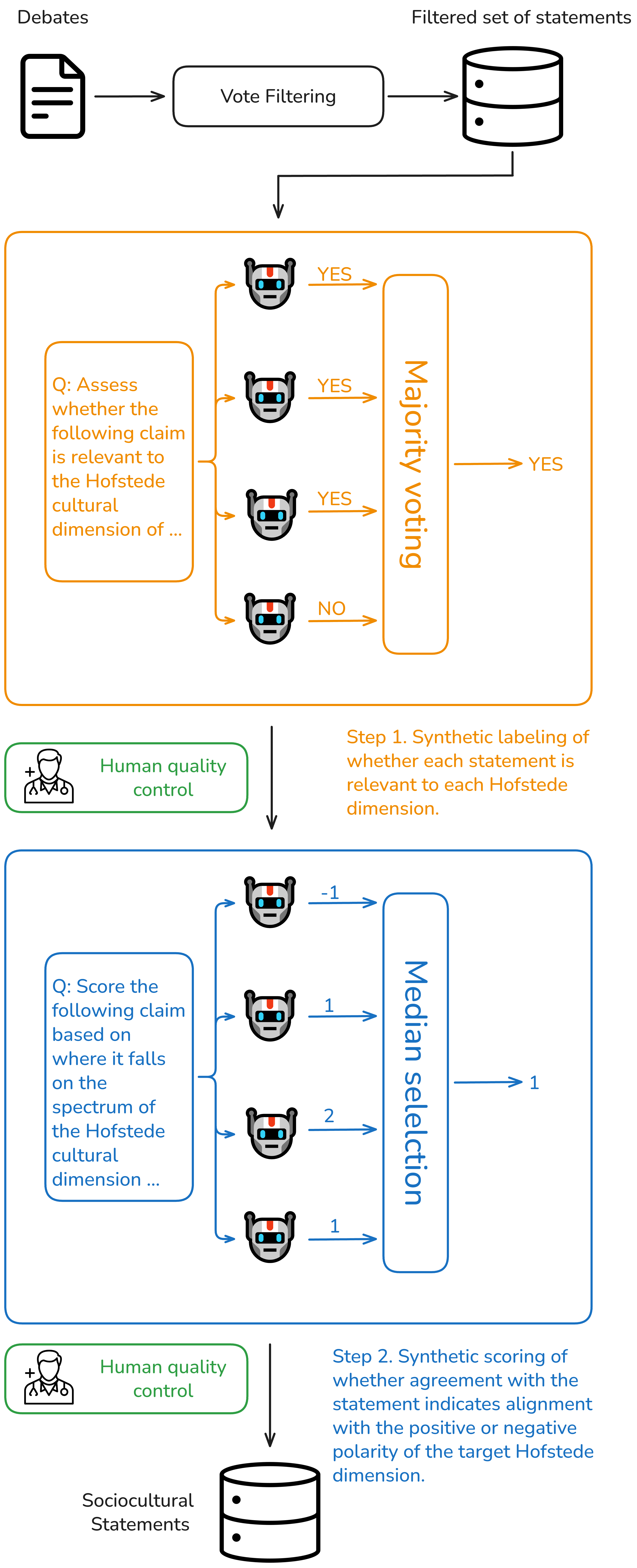}
    \vspace{-1mm}
    \caption{The statement scoring pipeline based on the six Hofstede dimensions. We collect natural debate statements via web scraping and filter them based on vote counts. Step 1 further removes statements that are irrelevant to the Hofstede dimensions. Step 2 assigns a score ($-2$ to $+2$), indicating how agreement with a statement aligns with the negative or positive polarity of each Hofstede dimension. Each synthetic labeling step is manually verified on a representative sample.}
    \vspace{-4mm}
    \label{fig:architecture}
\end{figure}

We propose a multi-stage methodology for scoring LLMs along the six Hofstede cultural dimensions. First, we introduce a {\em statement} dataset, consisting of topical claims that are potentially aligned with one or more of these dimensions. For example, supporting the statement {\em ``Countries should invest in renewable energies''} indicates strong alignment with the Long Term Orientation (LTO) dimension.
We collect these statements via web scraping from Kialo, an online debate platform.\footref{kialourl} We then assign synthetic dimension labels through a multi-step annotation pipeline. These labels provide scores that can measure the alignment of agreeing/disagreeing with a statement  
to a certain Hofstede dimension. \autoref{fig:architecture} describes the dataset generation process, further detailed below.

\subsection{Statement extraction and filtering}

In order to realistically assess alignment with cultural dimensions, statements should capture real-world, important, and complex societal issues. We found the Kialo platform to be a good source of such debates.\footref{kialourl} 
%
Kialo is a web platform designed for structured debate and argument analysis, consisting of thousands of individual debates on a broad area of subjects: politics, philosophy, science, ethics, religion. Each debate starts with a general statement (e.g., {\em ``Wealthy countries should provide citizens with a universal basic income''}), and users may provide arguments in favor or against that statement. A voting system ensures statement relevance and the strength of the arguments in the debate process.


We construct our dataset by first scraping these debates, from which we extract the initial statements and the number of votes. This vote count is then used as a filtering measure, to ensure the quality and relevance of the statement in our dataset. That is, since statements with more votes reflect a higher degree of user approval on their perceived reliability, we ensure that we filter out content that is potentially harmful or irrelevant. In particular, in this work, we disregard debates with a vote count of less than 3.
We include examples of the scraped statements in \autoref{sec:dataset_appendix}.

\subsection{Statement scoring}
The selected statements undergo a multi-step synthetic labeling process. In each step we aggregate the responses of several culturally-different LLMs (i.e., U.S. and Chinese), also prompting each in two different languages: English and Chinese.
We used GPT, Llama, DeepSeek and Qwen.

Each of these ``LLM-as-a-judge'' steps is quality-controlled by a team of human judges.
We use this multicultural, multilingual approach for scoring in order to account for cultural and linguistic variability, which allows us to obtain a cross-cultural interpretation of the annotated statements.

Our two-step labeling pipeline is guided by the following questions:
{\flushleft {\bf (1) Is a given statement relevant to a particular Hofstede dimension?}}

Scraped statements have a wide variety with respect to the societal issues they address. Moreover, as the Hofstede dimensions are not mutually exclusive, 
some statements may be relevant for more than one dimension, while other may not meaningfully align with any of them. For instance, agreement with the statement {\em ``Governments should make an effort to reduce the gender pay gap''} is informative for PDI and MAS, while being unrelated to the IND and UAI dimensions. 


To robustly annotate this non-trivial one-to-many alignment between statements and Hofstede dimensions, we decompose the annotation process into a series of binary questions, i.e., is statement $X$ relevant for dimension $Y$? LLMs are prompted, both in English and Chinese, to answer this question for each combination of statement and dimension in our dataset. We then use majority voting to aggregate the responses and keep only alignments that are relevant in at least half of the total number of predictions (i.e., at least 4 out of 8 total votes).


To ensure that majority voting between multiple LLMs yields correct annotations we performed a subsequent human quality control step. See ~\autoref{sec:humanqc} for details.

{\flushleft {\bf (2) Does agreeing with a statement indicate positive or negative polarity for a certain Hofstede dimension?}}

Next, we quantify where agreement with a specific claim falls on a discrete polarity scale for a given cultural dimension. In this work, we consider disagreement with a statement to have the opposite score as agreement.\footnote{Initial experiments with human annotators supported this assumption.}


We employ a discrete scoring system with the following possible values and interpretations (for agreement with the statement):
\begin{description}
    \vspace{-2mm}
    \item[~~~~$-2$:] strong negative polarity
    \vspace{-2mm}
    \item[~~~~$-1$:] slight negative polarity
    \vspace{-2mm}
    \item[~~~~$+1$:] slight positive polarity
    \vspace{-2mm}
    \item[~~~~$+2$:] strong positive polarity
    \vspace{-2mm}
\end{description}

Since statements are filtered for relevance in the previous step, a score of $0$ (neutral) is not permitted. 

Scores produced by the multicultural and multilingual set of LLMs are aggregated into a single statement-level score per relevant dimension by taking the median of the score distribution. We chose the median rather than the mean as it is less sensitive to outliers, and it reflects the central tendency of scores across culturally situated model outputs.

Similar to the previous annotation step, we perform human quality control here as well. 
 
\subsection{Human quality control}
\label{sec:humanqc}

We conduct a manual analysis to assess the quality of the generated labels, following both annotation steps described in the previous section. For this process, four human annotators manually evaluated the aggregated LLM responses.

This experiment was carried out across a subset of 56 statements, containing a total of 81 predictions, as some statements may be relevant for multiple different dimensions.

For the first annotation step, our manual quality control process focused on the {\em precision} of the LLM annotations. That is, the human annotators evaluated only the statements the LLMs considered relevant for a specific Hofstede dimension. The goal of this step is to ensure that the statement/dimension pairs that enter the second annotation step are as correct as possible.\footnote{Considering that our dataset is considerably larger than previous work in this space \cite{cao-etal-2023-assessing, culturefactor2024aimeetsculture}, we are less concerned about the existence of false negatives than false positives.} 
Across all evaluated instances, the human annotators confirmed the LLMs positive labels in 94.65\% of cases. We consider Cohen Kappa agreement as being irrelevant, due to this observed high agreement and the task being binary.
All in all, this analysis provides strong support for our LLM-as-a-judge setting for this annotation step.


\begin{table}[t!]

\begin{center}
\resizebox{1\linewidth
}{!}{
\begin{tabularx}{1.2\linewidth}{ l *{2}{Z} }
\hline
\textbf{Dim.} & \textbf{Inter-annotator pairwise average (MAE)}& \textbf{Annotators agg. vs LLMs agg. (MAE)} \\
\hline
PDI & 1.16 & 0.54 \\
IDV & 1.20 & 0.52 \\
MAS & 1.18 & 0.59\\
UAI & 0.54 & 0.50\\
LTO & 0.50 & 0.33   \\
IND & 0.69 & 0.58   \\
\hline
\end{tabularx}
}
\end{center}
\vspace{-3mm}
\caption{Inter-annotator pairwise average MAE (second column), and MAE between human annotators' and LLMs' aggregated scores (third column). Both MAE scores were measure for step 2 of our annotation framework. The aggregated value is the median of the scores.}
\vspace{-3mm}
\label{tab:MAE_scores}                                   
\end{table}

To evaluate the quality of the second step, the annotators independently scored the statements using the scoring system defined in the previous section. 
For this step, we measured: (a) inter-annotator disagreement using the mean absolute error (MAE) between all pairs of annotators on a $-2$ to $+2$ scoring scale; and (b) MAE between the aggregated human annotator score and the aggregated LLM score. To aggregate human scores we used the median of the distribution, similarly to what we did for LLMs. 
\autoref{tab:MAE_scores} shows the results of this analysis.


As the table indicates, 
across the six dimensions, the MAE between annotators ranged from $0.50$ to $1.20$, indicating non-trivial levels of disagreement. However, we consider this variation to be expected, given the heterogeneous composition of the annotator pool in terms of cultural background (i.e., U.S., E.U., Eastern Europe) and generational perspectives. Similar to \citet{xu2025noisesignalselbstzweckreframing}, who view annotator disagreement as a signal rather than noise, we treat label variation as a meaningful expression of diverse, and equally valid, perspectives.

To account for this diversity, we reported the MAE between the {\em aggregated} human score and the aggregated LLM one, using the medians of the respective distributions. 
Across all dimensions, the MAE between the LLM scores and the aggregated annotator scores ranged from $0.33$ to $0.59$, showing substantially smaller deviations from the aggregated annotator judgments. This indicates that the final scores generated by LLMs are generally well aligned with the central tendency of culturally diverse human judgments.




\section{Evaluation Methodology}

Our overall goal in this work is to measure the cultural biases that LLMs may have in response to prompts that test a variety of societal aspects. We use the statement dataset detailed in the previous section and the generated scores to compute the LLM biases along all Hofstede dimensions. Each evaluated LLM is prompted in both English and Chinese on whether it agrees or disagrees with the given dataset statement. This process is driven by a Chain-of-Thought prompt \cite{wei2023chainofthoughtpromptingelicitsreasoning}, which also asks the model to present the arguments that are for and against the given statement. Based on the final responses, we compute an average score per dimension, which tells us the model's overall position relative to each Hofstede dimension. 




\section{Experimental results}
\subsection{Experimental settings}

For the generation of the statements dataset and evaluation, we used 4 LLMs from 4 different families, two of which are developed in the United States, while the others are representative to China. The used models are GPT 4.1 (\texttt{gpt-4.1-2025-04-14}), Llama 3 (\texttt{llama-3-70b-instruct}), and DeepSeek-R1 (\texttt{deepseek-r1-0528}), and Qwen 2.5 (\texttt{qwen-2.5-72b-instruct}) respectively. Due to the variability in model responses throughout the entire labeling pipeline and evaluation, we selected the most frequent predictions from 5 different runs, using temperature 1.

\subsection{Results and discussion}
We present the scores of the four models when evaluated on the {\dataset} dataset. We use the empirically measured Hofstede values of the two nations\footnote{\label{hofstede_values}As measured by \url{https://www.theculturefactor.com}} (i.e., U.S. and China), derived from human measurements, as a comparison baseline. In the first two subsections we compare the models when prompted in their ``native'' languages (i.e., English for GPT and Llama, and Chinese for DeepSeek and Qwen). The third subsection presents the cultural differences that individual models exhibit across different languages.

\begin{figure*}[t]
    \begin{subfigure}[b]{0.32\textwidth}
        \centering
        \includegraphics[width=0.9\linewidth]{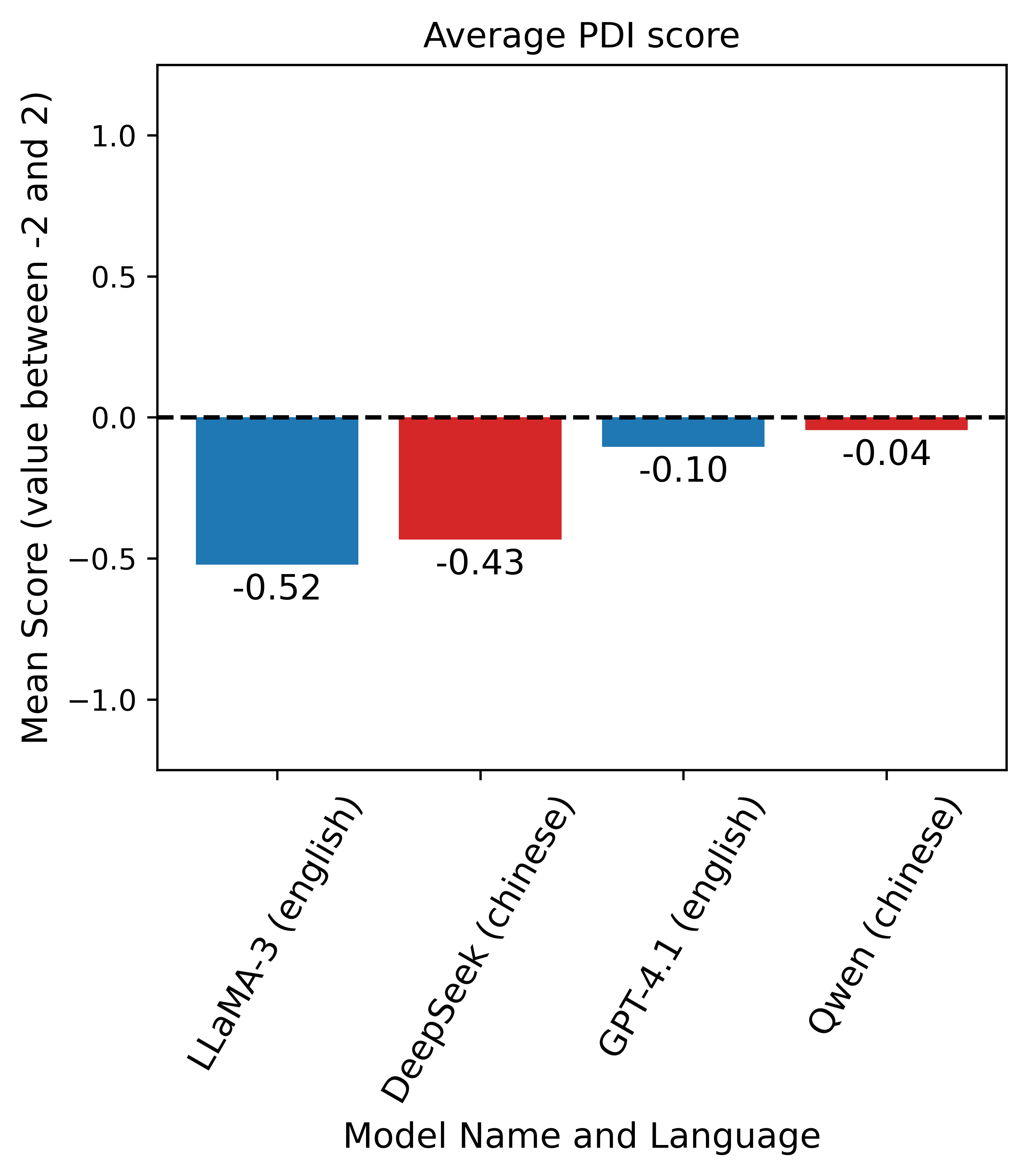}
        \subcaption{}
        \label{fig:PDI_all}
    \end{subfigure}
    \hfill
    \begin{subfigure}[b]{0.32\textwidth}
        \centering
        \includegraphics[width=0.9\linewidth]{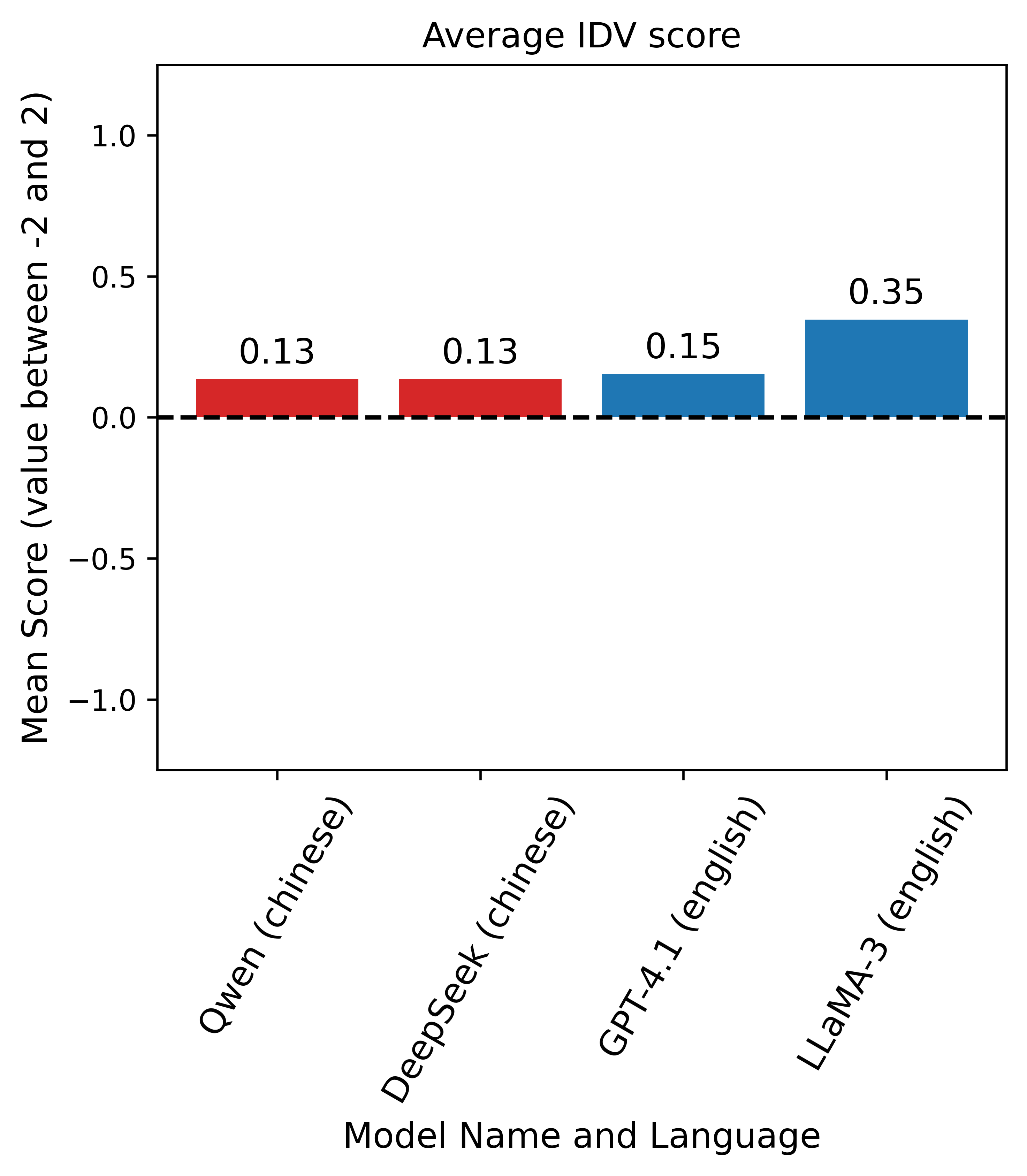}
        \subcaption{}
        \label{fig:IDV_all}
    \end{subfigure}
    \begin{subfigure}[b]{0.32\textwidth}
        \centering
        \includegraphics[width=0.9\linewidth]{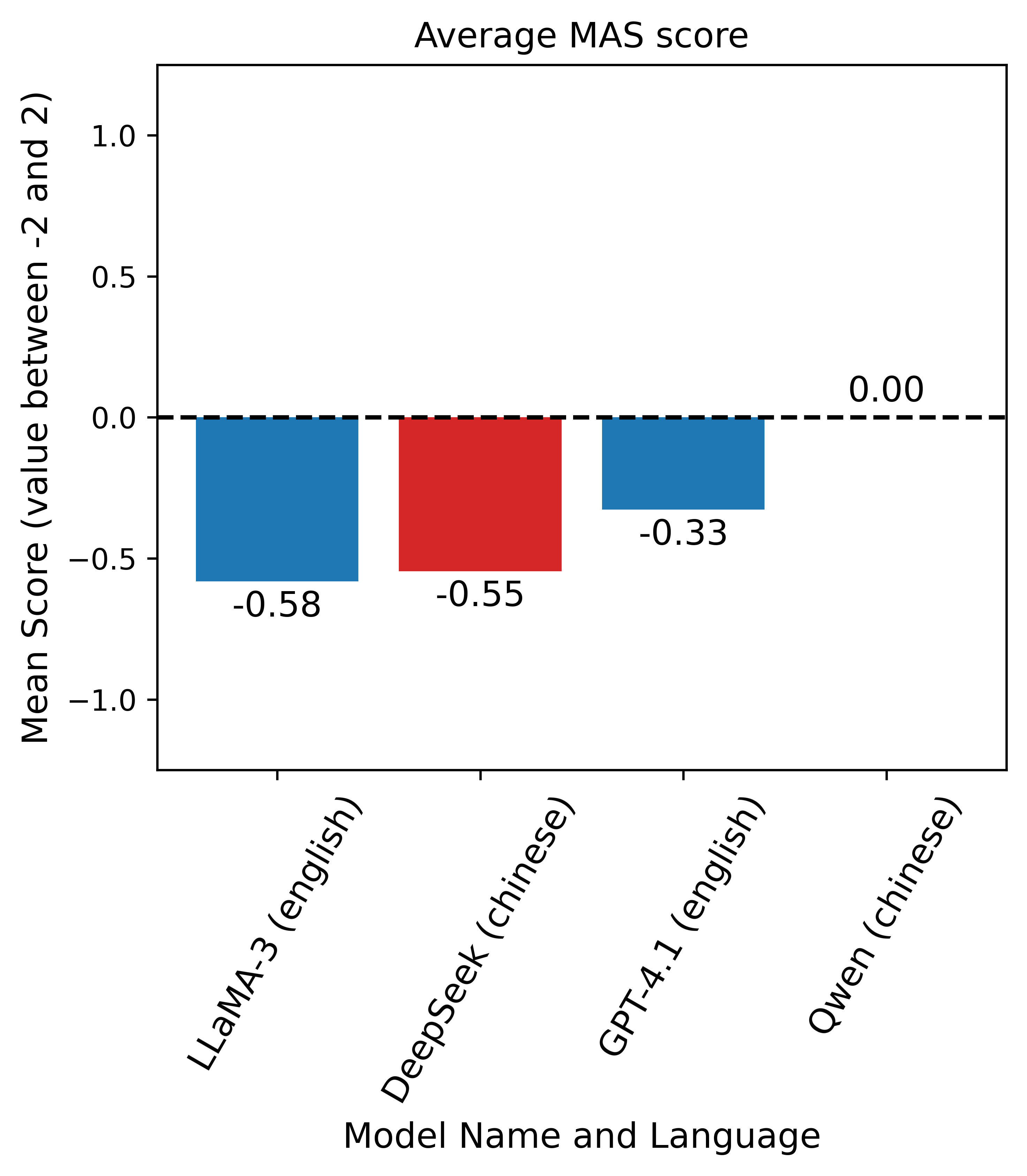}
        \subcaption{}
        \label{fig:MAS_all}
    \end{subfigure}
    \\
    \begin{subfigure}[b]{0.32\textwidth}
        \centering
        \includegraphics[width=0.9\linewidth]{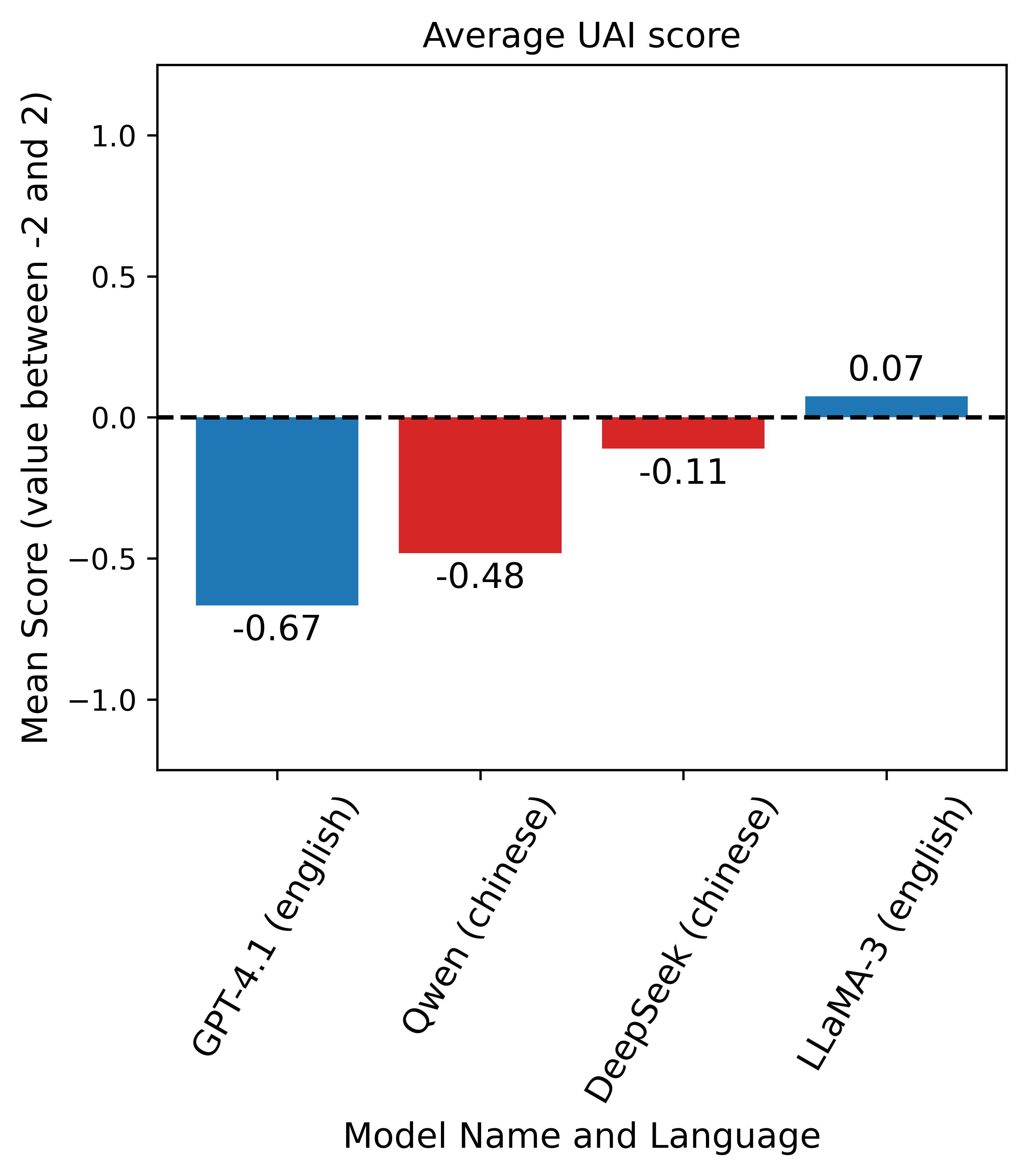}
        \subcaption{}
        \label{fig:UAI_all}
    \end{subfigure}
    \hfill
    \begin{subfigure}[b]{0.32\textwidth}
        \centering
        \includegraphics[width=0.9\linewidth]{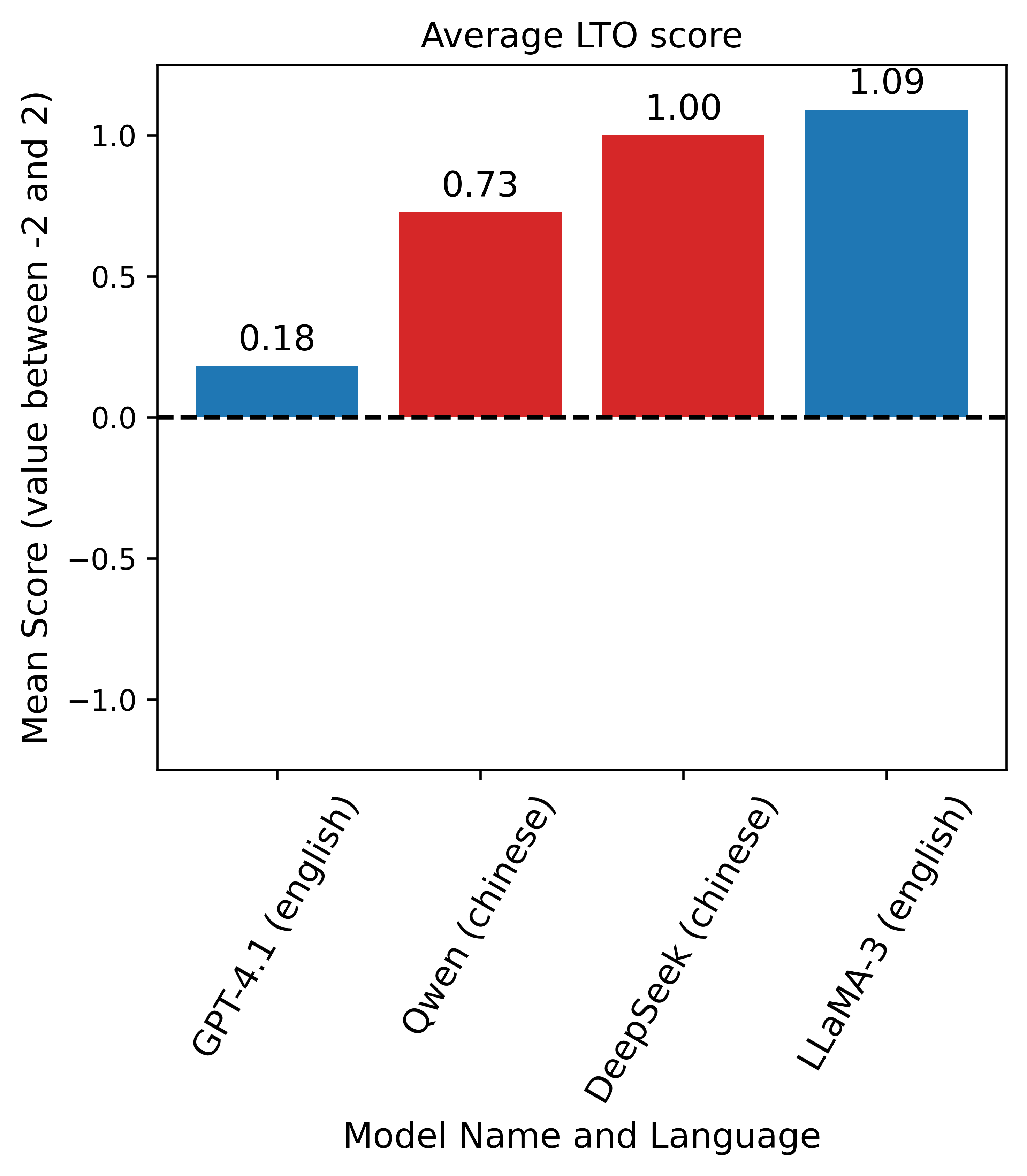}
        \subcaption{}
        \label{fig:LTO_all}
    \end{subfigure}
    \begin{subfigure}[b]{0.32\textwidth}
        \centering
        \includegraphics[width=0.9\linewidth]{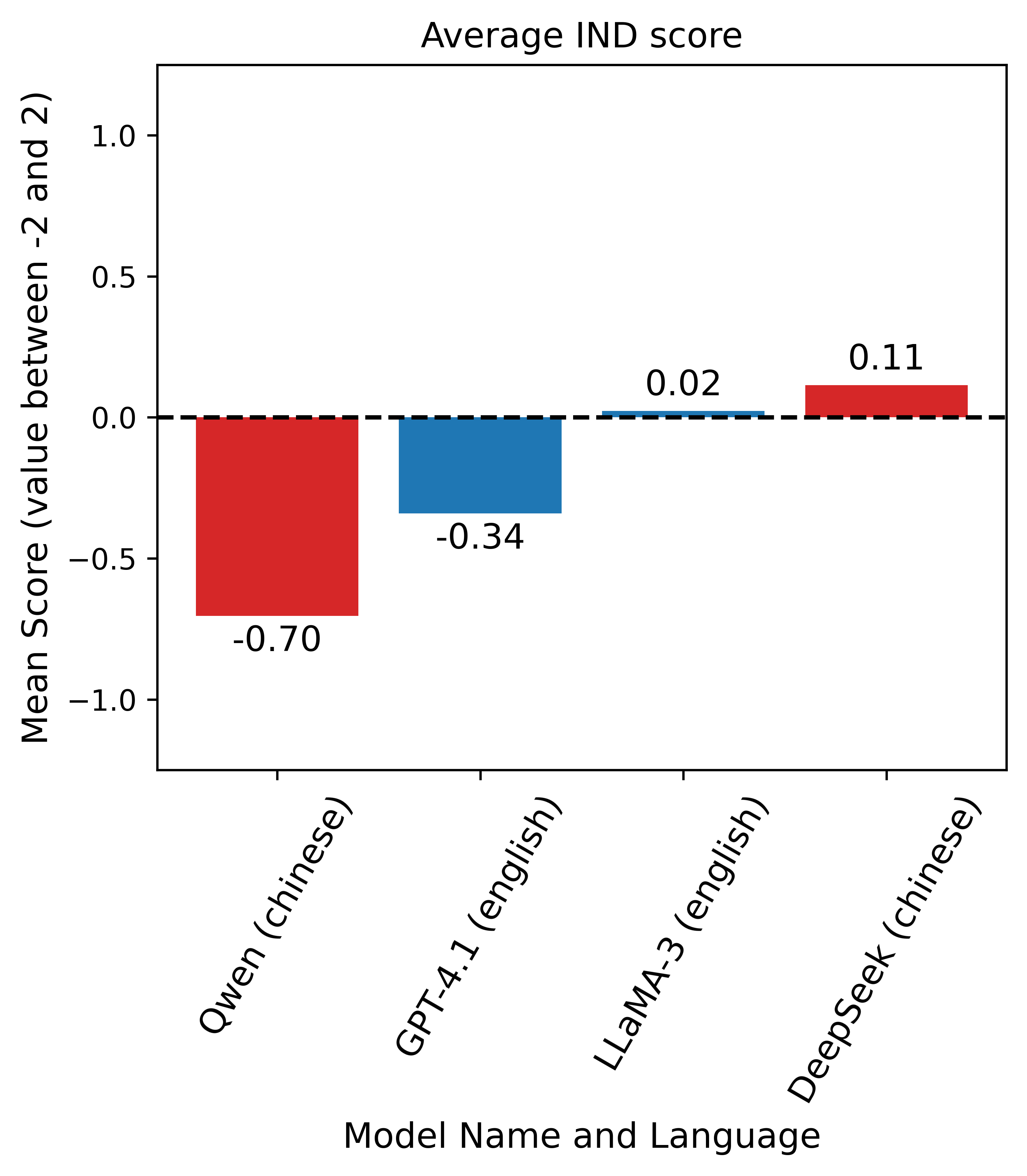}
        \subcaption{}
        \label{fig:IND_all}
    \end{subfigure}
    \vspace{-1mm}
    \caption{Average scores across all Hofstede dimensions for each model when prompted in its ''native`` language, evaluated on {\dataset} dataset. The red bars represent Chinese models, while the blue bars represent U.S. models. Despite values being in the interval [-2, 2], we used [-1, 1] for the axis as it fits the average scores.}
    \vspace{-4mm}
    \label{fig:all_results}
\end{figure*}

\subsubsection{Comparing U.S. vs. Chinese LLMs}
\label{subsec:llm_groups}

 \autoref{fig:english_vs_chinese} shows a comparison between the aggregated responses from the Chinese models and the U.S. models, when prompted in their ``native'' languages. The aggregated response is the median of two LLMs per culture, where each LLM was run 5 times.
 We compare the aggregated LLM responses with the actual (human) Hofstede values of the two nations\footref{hofstede_values}. To highlight the sociocultural {\em differences} between the two LLM groups, we considered only the statements where the two groups responded differently for this experiment. To align the scale of our experiments ($-2$ to $2$) with the scale of the measurements for the human sociocultural differences ($0$ to $100$), we then applied min-max scaling. 
 
 As the figure shows, {\em culturally-distinct LLM groups follow trends similar to those observed among people from the same culture across all Hofstede dimensions}.
  For example, a higher PDI that is present for both Chinese human measurements and Chinese LLMs shows an acceptance for hierarchical structures and authority. In contrast, Western cultures with a lower PDI favor more egalitarian values. This is illustrated by the statement {\em ``Unpaid internships should be banned''}, with which U.S. LLMs generally agree, while Chinese models tend to disagree. To detail, the U.S. LLMs consider unpaid internships as unequal treatment (thus, showing a low PDI), while the Chinese LLMs accept hierarchical differences and see unpaid internships as a privilege granted by those with power (i.e., higher PDI). 
  

  
 The most notable difference is observed for the IND metric. For example, the two LLM groups had contradictory predictions for the statement {\em``Developed countries should reduce the working year.''}; the U.S. models agreed to the statement, while Chinese models rejected it. 
 This difference highlights that the U.S. models reflect a sociocultural tendency toward indulgence (as defined by Hofstede), whereas the Chinese models align more with values of restraint. See \autoref{appendix:examples_appendix} for more examples.
 
 Overall, this comparison indicates that the models do, in fact, reflect the sociocultural values of the countries in which they were developed. This is an important observation, given that these models are designed for international use, where many users come from cultures with different values and norms. This disconnect strongly suggest a pragmatic conclusion: a single builtin alignment model is not sufficient; we need LLMs that can accommodate multiple forms of alignment, each suited to the sociocultural context of their target users.
 


\subsubsection{Comparing individual models on their ``native'' languages}

\autoref{fig:all_results} compares the models used in our experiments when prompted in their respective ``native'' languages. In this experiment, we aggregate the scores from all statements relevant to each dimension, rather than focusing solely on the differences (as in the previous subsection).

We observe several deviations from the trends established by human measurements. For example, in  \autoref{fig:PDI_all}, DeepSeek’s PDI values are low and closely resemble those of the U.S. models. This contradicts the sociocultural values observed in contemporary China \cite{Hofstede:1991}, where hierarchy is generally accepted and PDI scores are consistently high. Another  outlier appears in \autoref{fig:MAS_all}, where DeepSeek shows a low MAS score, in contrast to the high score associated with Chinese culture, which reflects goal-oriented behavior and a clearer delineation of gender roles. DeepSeek also diverges from cultural expectations for IND, as shown in \autoref{fig:IND_all}, where it exhibits the highest level of indulgence among the models. Overall, these results indicate a misalignment between DeepSeek’s predictions and the Hofstede values associated with Chinese culture.
In contrast, Qwen’s predictions closely align with the sociocultural values of China across all dimensions.

There are multiple possible interpretations for the observed differences between Qwen and DeepSeek. 
For example, \citet{Jin2023NewChinaPlaybook} presents the Chinese economic, social, and political model as complex and difficult to categorize on the spectrum between socialism and capitalism. The work characterizes Chinese culture as a heterogeneous hybrid system that combines socialism with Confucian principles (for the social fabric of the Chinese society), while also valuing traits typically associated with capitalist systems such as ``fierce'' competition and meritocracy (as the engine driving the country's economical progress).

\citet{Jin2023NewChinaPlaybook} suggests that Western sociocultural categorization systems (e.g., Hofstede dimensions) should be context-dependent (e.g., along authority, family, market, or ideology axes). 
For example, Qwen’s high PDI score may reflect Confucian values that remain prevalent in modern China in the context of of family or formal authority. In contrast, a low PDI score could indicate the influence of socialist ideology, which emphasizes equality. Similar observations can be made for the other dimensions as well.

Considering these observations, it is likely that the two Chinese models capture different, yet equally valid, perspectives of Chinese culture. A more mundane explanation is also possible: DeepSeek may have been developed with a more international user base in mind, where Western sociocultural values are perceived as dominant. Nevertheless, in aggregate, the two Chinese models closely reflect Chinese sociocultural values.

Similar patterns are observed for the U.S. models. Llama exhibits sociocultural values that are somewhat more left-leaning compared to GPT. However, when considered in aggregate, both models align well with U.S. sociocultural values.



\subsubsection{Comparing individual models across different languages}
\autoref{fig:language_comparison} shows the average scores across all Hofstede dimensions for each model, when prompted in English and Chinese. The cells in green present cases where the cross-lingual sociocultural differences are statistically significant ($p$-values $< 0.05$ under a bootstrap resampling significance test). 

This experiment enables us to assess whether changing the prompt language influences the sociocultural values expressed by the model.
We observe moderate cross-lingual differences within individual models, particularly for DeepSeek and GPT, where more statistically significant differences appear: for GPT in UAI and LTO, and for DeepSeek in PDI and LTO. We consider this phenomenon normal, as it may indicate that the models are somewhat differently aligned for the two languages and the respective cultures.

In contrast, it is potentially concerning that for most other dimensions and models, no statistically significant differences are observed. This suggests that these models may culturally align their outputs to a common profile, regardless of differences in users’ sociocultural backgrounds.



\begin{table}[h!]
\begin{center}
\resizebox{1\linewidth
}{!}{
\begin{tabularx}{1.2\linewidth}{ c l *{4}{Z} }
\hline
\textbf{Dim.} & \textbf{Lang.} & \textbf{GPT} & \textbf{Llama} & \textbf{DeepSeek} & \textbf{Qwen} \\
\hline

\multirow{2}{*}{{PDI}}
  & English  & -0.10   & -0.52  & \cellcolor{lightgreen}-0.22  & -0.07  \\
  & Chinese  & -0.25 & -0.31 & \cellcolor{lightgreen}-0.43 & -0.04 \\
\hline

\multirow{2}{*}{{IDV}}
  & English  & 0.15  & 0.35  & 0.19  & 0.04  \\
  & Chinese  & 0.12 & 0.25 & 0.13 & 0.13 \\
\hline

\multirow{2}{*}{{MAS}}
  & English  & -0.33  & -0.58 & -0.65  & 0.0 \\
  & Chinese  & -0.44 & -0.44 & -0.55 & 0.0 \\
\hline

\multirow{2}{*}{{UAI}}
  & English  & \cellcolor{lightgreen}-0.67  & \cellcolor{lightgreen}0.07  & -0.11  & \cellcolor{lightgreen}-0.74  \\
  & Chinese  & \cellcolor{lightgreen}-0.41 & \cellcolor{lightgreen}-0.3 & -0.11 & \cellcolor{lightgreen}-0.48 \\
\hline

\multirow{2}{*}{{LTO}}
  & English  & \cellcolor{lightgreen}0.18  & 1.09   & \cellcolor{lightgreen}1.09  & 0.18  \\
  & Chinese  & \cellcolor{lightgreen}0.64 & 1.18 & \cellcolor{lightgreen}1.00 & 0.73 \\
\hline

\multirow{2}{*}{{IND}}
  & English  & -0.34 & 0.02   & -0.02  & -0.61 \\
  & Chinese  & -0.02 & 0.25  &  0.11 & -0.7 \\
\hline
\end{tabularx}
}
\end{center}
\vspace{-3mm}
\caption{Average scores across all Hofstede dimensions for each model when prompted in both languages, evaluated on {\dataset} dataset. The scores represent inclinations towards the positive or negative polarity of the six dimensions. The green highlighting indicates statistically significant differences.}
\vspace{-5mm}
\label{fig:language_comparison}
\end{table}

\subsection{Potential explanations}

Our analysis shows that LLMs reflect the sociocultural values of their creators. This alignment may arise from two main sources: pre-training and instruction tuning/alignment. During pre-training, it is possible that model developers bias the web-crawled data toward their own language or sociocultural values. However, given the scale and diversity of the web, it is unlikely that pre-training datasets are deliberately manipulated in a substantial way.

A more plausible explanation is that sociocultural values are introduced during the instruction-tuning and alignment phase, which remains largely opaque. If this is the case, our findings underscore the importance of greater transparency in instruction tuning, including openness about the instruction datasets used and the training procedures.



\section{Conclusions}

We investigated how LLMs encode the sociocultural contexts of their human creators.
We introduced {\dataset}, a dataset constructed from natural debate statements using a multi-step methodology. The dataset includes synthetic labels that quantify the Hofstede sociocultural dimensions reflected in LLM responses to these statements. We conducted a comparative multilingual analysis of two groups of LLMs developed in different countries (U.S. and China) and compared the results with human sociocultural values.

Our results indicate that LLMs reflect the sociocultural values and norms of the countries in which they were developed. Moreover, in general, LLMs do not exhibit statistically significant differences in sociocultural values when the prompt language is changed. 
These findings underscore the limited ability of current LLMs to adapt to the sociocultural backgrounds of their international users. This limitation should be a key consideration in the development of future LLMs, particularly during the alignment process. We anticipate that our study will contribute to the development of multiple culturally informed alignment strategies, each tailored to the cultural context of the target users.

\section*{Limitations}


Our work quantifies and analyzes the sociocultural values expressed by LLMs developed in two culturally distinct regions (the United States and China), using the Hofstede dimensions as a reference framework. Our analysis is limited to two models per country when constructing aggregated predictions. This constraint is driven by the large number of model queries required for both dataset construction and evaluation.

Specifically, dataset annotation was performed using all four analyzed LLMs, each prompted in two languages (English and Chinese), with five runs per model–language combination. Starting from approximately 1,500 statements, this process resulted in roughly 100,000 model requests across dataset creation and analysis. Expanding the study to include additional cultural regions or a larger number of models per region would substantially increase the number of required requests and the overall cost.

Lastly, the Chinese translations of the prompts and statements were performed using Google Translate and were not validated by a Chinese speaker.


\section*{Ethics}


This work analyzes the sociocultural values expressed by LLMs when responding to real-world debate statements collected from an online platform. Because these debates address complex societal issues, they may include sensitive or potentially offensive content, such as references to drugs, sexuality, religion, or war. We intentionally did not filter out such topics in order to preserve realism and to measure the sociocultural values expressed by LLMs as faithfully as possible. As the dataset is used solely for analytical purposes and the models are employed exclusively for inference, this work neither introduces new harmful content nor modifies the models in any way that could increase harm.

We use these cultures to identify sociocultural biases in LLMs. It is not our intention to compare cultures. 


\bibliography{custom}

@article{yang-etal-2025-socially,
    title = "Socially Aware Language Technologies: Perspectives and Practices",
    author = "Yang, Diyi  and
      Hovy, Dirk  and
      Jurgens, David  and
      Plank, Barbara",
    journal = "Computational Linguistics",
    volume = "51",
    month = jun,
    year = "2025",
    address = "Cambridge, MA",
    publisher = "MIT Press",
    url = "https://aclanthology.org/2025.cl-2.10/",
    doi = "10.1162/coli_a_00556",
    pages = "689--703"
}

@misc{kharchenko2025llmsrepresentvaluescultures,
      title={How Well Do LLMs Represent Values Across Cultures? Empirical Analysis of LLM Responses Based on Hofstede Cultural Dimensions}, 
      author={Julia Kharchenko and Tanya Roosta and Aman Chadha and Chirag Shah},
      year={2025},
      eprint={2406.14805},
      archivePrefix={arXiv},
      primaryClass={cs.CL},
      url={https://arxiv.org/abs/2406.14805}, 
}

@inproceedings{masoud-etal-2025-cultural,
    title = "Cultural Alignment in Large Language Models: An Explanatory Analysis Based on Hofstede{'}s Cultural Dimensions",
    author = "Masoud, Reem  and
      Liu, Ziquan  and
      Ferianc, Martin  and
      Treleaven, Philip C.  and
      Rodrigues, Miguel Rodrigues",
    editor = "Rambow, Owen  and
      Wanner, Leo  and
      Apidianaki, Marianna  and
      Al-Khalifa, Hend  and
      Eugenio, Barbara Di  and
      Schockaert, Steven",
    booktitle = "Proceedings of the 31st International Conference on Computational Linguistics",
    month = jan,
    year = "2025",
    address = "Abu Dhabi, UAE",
    publisher = "Association for Computational Linguistics",
    url = "https://aclanthology.org/2025.coling-main.567/",
    pages = "8474--8503"
}

@misc{ma2025openopinionembracingopenendedness,
      title={Too Open for Opinion? Embracing Open-Endedness in Large Language Models for Social Simulation}, 
      author={Bolei Ma and Yong Cao and Indira Sen and Anna-Carolina Haensch and Frauke Kreuter and Barbara Plank and Daniel Hershcovich},
      year={2025},
      eprint={2510.13884},
      archivePrefix={arXiv},
      primaryClass={cs.CL},
      url={https://arxiv.org/abs/2510.13884}, 
}

@misc{xu2025noisesignalselbstzweckreframing,
      title={From Noise to Signal to Selbstzweck: Reframing Human Label Variation in the Era of Post-training in NLP}, 
      author={Shanshan Xu and Santosh T. Y. S. S and Barbara Plank},
      year={2025},
      eprint={2510.12817},
      archivePrefix={arXiv},
      primaryClass={cs.CL},
      url={https://arxiv.org/abs/2510.12817}, 
}

@misc{wei2023chainofthoughtpromptingelicitsreasoning,
      title={Chain-of-Thought Prompting Elicits Reasoning in Large Language Models}, 
      author={Jason Wei and Xuezhi Wang and Dale Schuurmans and Maarten Bosma and Brian Ichter and Fei Xia and Ed Chi and Quoc Le and Denny Zhou},
      year={2023},
      eprint={2201.11903},
      archivePrefix={arXiv},
      primaryClass={cs.CL},
      url={https://arxiv.org/abs/2201.11903}, 
}

@misc{culturefactor2024aimeetsculture,
  title        = {AI Meets Culture: Bridging the Divide Between Technology and Human Values},
  author       = {{The Culture Factor}},
  year         = {2024},
  howpublished = {\url{https://news.theculturefactor.com/news/ai-meets-culture}},
  note         = {Accessed: 2025-03-08}
}

@book{Hofstede:1991,
  address = {London and New York},
  author = {Hofstede, Geert H.},
  booktitle = {Cultures and Organizations: Software of the Mind},
  pages = {xii, 279},
  publisher = {McGraw-Hill},
  title = {Cultures and organizations: Software of the mind},
  year = 1991
}

@inproceedings{shen-etal-2024-understanding,
    title = "Understanding the Capabilities and Limitations of Large Language Models for Cultural Commonsense",
    author = "Shen, Siqi  and
      Logeswaran, Lajanugen  and
      Lee, Moontae  and
      Lee, Honglak  and
      Poria, Soujanya  and
      Mihalcea, Rada",
    editor = "Duh, Kevin  and
      Gomez, Helena  and
      Bethard, Steven",
    booktitle = "Proceedings of the 2024 Conference of the North American Chapter of the Association for Computational Linguistics: Human Language Technologies (Volume 1: Long Papers)",
    month = jun,
    year = "2024",
    address = "Mexico City, Mexico",
    publisher = "Association for Computational Linguistics",
    url = "https://aclanthology.org/2024.naacl-long.316/",
    doi = "10.18653/v1/2024.naacl-long.316",
    pages = "5668--5680"
}

@book{Jin2023NewChinaPlaybook,
  author    = "Jin, Keyu",
  title     = "The New China Playbook: Beyond Socialism and Capitalism",
  publisher = "Viking",
  address   = "New York",
  year      = "2023",
  isbn      = "9781984878298",
}

@techreport{chatgptUse,
 title = "How People Use ChatGPT",
 author = "Chatterji, Aaron and Cunningham, Thomas and Deming, David J and Hitzig, Zoe and Ong, Christopher and Shan, Carl Yan and Wadman, Kevin",
 institution = "National Bureau of Economic Research",
 type = "Working Paper",
 series = "Working Paper Series",
 number = "34255",
 year = "2025",
 month = "September",
 doi = {10.3386/w34255},
 URL = "http://www.nber.org/papers/w34255",
}

@inproceedings{AIWeird,
author = {Mihalcea, Rada and Ignat, Oana and Bai, Longju and Borah, Angana and Chiruzzo, Luis and Jin, Zhijing and Kwizera, Claude and Nwatu, Joan and Poria, Soujanya and Solorio, Thamar},
title = {Why AI is WEIRD and shouldn't be this way: towards AI for everyone, with everyone, by everyone},
year = {2025},
isbn = {978-1-57735-897-8},
publisher = {AAAI Press},
url = {https://doi.org/10.1609/aaai.v39i27.35092},
doi = {10.1609/aaai.v39i27.35092},
booktitle = {Proceedings of the Thirty-Ninth AAAI Conference on Artificial Intelligence and Thirty-Seventh Conference on Innovative Applications of Artificial Intelligence and Fifteenth Symposium on Educational Advances in Artificial Intelligence},
articleno = {3194},
numpages = {14},
series = {AAAI'25/IAAI'25/EAAI'25}
}

@article{10.1162/COLI.a.14,
    author = {Pawar, Siddhesh and Park, Junyeong and Jin, Jiho and Arora, Arnav and Myung, Junho and Yadav, Srishti and Haznitrama, Faiz Ghifari and Song, Inhwa and Oh, Alice and Augenstein, Isabelle},
    title = {Survey of Cultural Awareness in Language Models: Text and Beyond},
    journal = {Computational Linguistics},
    volume = {51},
    number = {3},
    pages = {907-1004},
    year = {2025},
    month = {09},
    issn = {0891-2017},
    doi = {10.1162/COLI.a.14},
    url = {https://doi.org/10.1162/COLI.a.14},
    eprint = {https://direct.mit.edu/coli/article-pdf/51/3/907/2523159/coli.a.14.pdf},
}

@misc{johnson2022ghostmachineamericanaccent,
      title={The Ghost in the Machine has an American accent: value conflict in GPT-3}, 
      author={Rebecca L Johnson and Giada Pistilli and Natalia Menédez-González and Leslye Denisse Dias Duran and Enrico Panai and Julija Kalpokiene and Donald Jay Bertulfo},
      year={2022},
      eprint={2203.07785},
      archivePrefix={arXiv},
      primaryClass={cs.CL},
      url={https://arxiv.org/abs/2203.07785}, 
}

@inproceedings{Jakesch_2023, series={CHI ’23},
   title={Co-Writing with Opinionated Language Models Affects Users’ Views},
   url={http://dx.doi.org/10.1145/3544548.3581196},
   DOI={10.1145/3544548.3581196},
   booktitle={Proceedings of the 2023 CHI Conference on Human Factors in Computing Systems},
   publisher={ACM},
   author={Jakesch, Maurice and Bhat, Advait and Buschek, Daniel and Zalmanson, Lior and Naaman, Mor},
   year={2023},
   month=apr, pages={1–15},
   collection={CHI ’23} }

@inproceedings{chawla-etal-2023-social,
    title = "Social Influence Dialogue Systems: A Survey of Datasets and Models For Social Influence Tasks",
    author = "Chawla, Kushal  and
      Shi, Weiyan  and
      Zhang, Jingwen  and
      Lucas, Gale  and
      Yu, Zhou  and
      Gratch, Jonathan",
    editor = "Vlachos, Andreas  and
      Augenstein, Isabelle",
    booktitle = "Proceedings of the 17th Conference of the European Chapter of the Association for Computational Linguistics",
    month = may,
    year = "2023",
    address = "Dubrovnik, Croatia",
    publisher = "Association for Computational Linguistics",
    url = "https://aclanthology.org/2023.eacl-main.53/",
    doi = "10.18653/v1/2023.eacl-main.53",
    pages = "750--766"
}

@inproceedings{plank-2022-problem,
    title = "The ``Problem'' of Human Label Variation: On Ground Truth in Data, Modeling and Evaluation",
    author = "Plank, Barbara",
    editor = "Goldberg, Yoav  and
      Kozareva, Zornitsa  and
      Zhang, Yue",
    booktitle = "Proceedings of the 2022 Conference on Empirical Methods in Natural Language Processing",
    month = dec,
    year = "2022",
    address = "Abu Dhabi, United Arab Emirates",
    publisher = "Association for Computational Linguistics",
    url = "https://aclanthology.org/2022.emnlp-main.731/",
    doi = "10.18653/v1/2022.emnlp-main.731",
    pages = "10671--10682"
}

@inproceedings{cao-etal-2023-assessing,
    title = "Assessing Cross-Cultural Alignment between {C}hat{GPT} and Human Societies: An Empirical Study",
    author = "Cao, Yong  and
      Zhou, Li  and
      Lee, Seolhwa  and
      Cabello, Laura  and
      Chen, Min  and
      Hershcovich, Daniel",
    editor = "Dev, Sunipa  and
      Prabhakaran, Vinodkumar  and
      Adelani, David Ifeoluwa  and
      Hovy, Dirk  and
      Benotti, Luciana",
    booktitle = "Proceedings of the First Workshop on Cross-Cultural Considerations in NLP (C3NLP)",
    month = may,
    year = "2023",
    address = "Dubrovnik, Croatia",
    publisher = "Association for Computational Linguistics",
    url = "https://aclanthology.org/2023.c3nlp-1.7/",
    doi = "10.18653/v1/2023.c3nlp-1.7",
    pages = "53--67"
}

@misc{prabhakaran2022culturalincongruenciesartificialintelligence,
      title={Cultural Incongruencies in Artificial Intelligence}, 
      author={Vinodkumar Prabhakaran and Rida Qadri and Ben Hutchinson},
      year={2022},
      eprint={2211.13069},
      archivePrefix={arXiv},
      primaryClass={cs.CY},
      url={https://arxiv.org/abs/2211.13069}, 
}
\newpage
\appendix

\section{Dataset statistics and examples}
\label{sec:dataset_appendix}

We started with a set of 1,534 scraped debates, 776 of which remain after the automated vote-count filtering process. After the first annotation step, we identified 68 statements that are relevant for PDI, 116 for IDV, 60 for MAS, 62 for UAI, 24 for LTO, and 48 for IND. \autoref{tab:claim_examples} offers examples of relevant statements for each dimension, associated with the assigned score. Even though this dataset may be considered small in the age of LLMs, it is considerably larger than existing work in this space \cite{cao-etal-2023-assessing, culturefactor2024aimeetsculture}.

We release the \dataset\ dataset publicly - \url{https://anonymous.4open.science/r/LLMs-Cultural-Values-E7EC/README.md}. 
Due to licensing issues, we release statement ids (instead of the statement texts) together with scrapping software, which guarantees that our work is reproducible.

\begin{table}[h!]
\begin{center}
\resizebox{1\linewidth
}{!}{
\begin{tabularx}{1.2\linewidth}{ J r r }
\hline
\textbf{Statement} & \textbf{Dim.} & \textbf{Score} \\
\hline
{\em The Electoral College should be abolished.} & PDI & -1 \\ 
\hline
{\em Organ donation should be mandatory.} & IDV & -2 \\ 
\hline
{\em The primary focus of prisons should be rehabilitation and reintegration, not punishment.} & MAS & -1 \\ 
\hline
{\em Health care providers should be mandated to get the flu vaccine.} & UAI & 1 \\
\hline
{\em Governments should push to implement 100\% renewable energy.} & LTO & 2 \\
\hline
{\em Marijuana should be as legally accessible as alcohol.} & IND & 2   \\

\hline
\end{tabularx}
}
\end{center}
\vspace{-3mm}
\caption{Examples of statements that are relevant for each Hofstede dimension, with their assigned score. In case of agreement with the statement, a positive score shows alignment with the positive polarity of the listed dimension, while a negative score shows alignment with the negative polarity of the dimension. Note that a statement may be relevant for multiple dimensions.} 
\vspace{-3mm}
\label{tab:claim_examples}
\end{table}

\section{Examples of sociocultural statements and LLM responses}
\label{appendix:examples_appendix}
In this section we analyze multiple statements, for different Hofstede dimensions, where the two LLM groups (U.S. and Chinese) responded differently.  

A statement where the two groups responded differently is {\em``Every human should have the right and means to decide when and how to die''}. The U.S. LLMs’ agreement reflects a more individualistic outlook, whereas the Chinese LLMs’ disagreement aligns with the lower IDV score associated with Chinese culture, in which individual decisions are expected to consider the welfare of one’s group and broader society.

For the statement {\em ``The UK should remain in the EU if a hard Brexit is the only alternative option''}, the U.S. models’ agreement aligns with a higher UAI, emphasizing stability and predictability, whereas the Chinese models’ disagreement indicates a greater tolerance for uncertainty in pursuit of potential long-term benefits.

\end{document}